\documentclass[useAMS,usenatbib]{mn2e}
\usepackage{times}
\usepackage{graphics,epsfig}
\usepackage{graphicx}
\usepackage{amssymb}

\title[Turbulence in the atomic ISM]{Turbulence measurements from H~{\sc i} absorption spectra}

\author[N. Roy, L. Peedikakkandy and J. N. Chengalur]{
Nirupam Roy $^{1}$\thanks{E-mail: nirupam@ncra.tifr.res.in~(NR); 
leshma2004@gmail.com~(LP); chengalu@ncra.tifr.res.in~(JNC);}, 
Leshma Peedikakkandy $^{2}$\footnotemark[1] and
Jayaram N. Chengalur $^{1}$\footnotemark[1]\\
       $^{1}$NCRA-TIFR, Post Bag 3, Ganeshkhind, Pune 411 007, India\\
       $^{2}$Department of Physics, Cochin University of Science \& Technology, 
Kochi 682 022, India}

\begin{document}
\date{Accepted yyyy month dd. Received yyyy month dd; in original form yyyy 
month dd}

\pagerange{\pageref{firstpage}--\pageref{lastpage}} \pubyear{2008}

\maketitle

\label{firstpage}

\begin{abstract}
We use the millennium Arecibo 21 cm absorption-line survey measurements to 
examine the issue of the non-thermal contribution to the observed Galactic 
H~{\sc i} line widths. If we assume a simple, constant pressure model for the 
H~{\sc i} in the Galaxy, we find that the non-thermal contribution to the line 
width, $v_{nt}$ scales as $v^2_{nt} \propto l^{\alpha}$, for $v_{nt}$ larger 
than $\sim 0.7$ km s$^{-1}$. Here $l$ is a derived length scale and $\alpha 
\sim 0.7 \pm 0.1$. This is consistent with what one would expect from a 
turbulent medium with a Kolmogorov scaling. Such a scaling is also predicted 
by theoretical models and numerical simulations of turbulence in a magnetized 
medium. For non-thermal line widths narrower than $\sim 0.7$ km s$^{-1}$, this 
scaling breaks down, and we find that the likely reason is ambiguities arising 
from Gaussian decomposition of intrinsically narrow, blended lines. We use the 
above estimate of the non-thermal contribution to the line width to determine 
corrected H~{\sc i} kinetic temperature. The new limits that we obtain imply 
that a significantly smaller ($\sim 40\%$ as opposed to 60\%) fraction of the 
atomic interstellar medium in our Galaxy is in the warm neutral medium phase. 
\end{abstract}


\section{Introduction}
\label{sec:int}

The classical description of the Galactic atomic interstellar medium (ISM) is 
that it consists of the cold neutral medium (CNM) and the warm neutral medium 
(WNM), in rough pressure balance with each other \citep[e.g.][]{fi65,fi69,wo03}.
Detailed  modeling of the energy balance in a multi-phase medium 
\citep[e.g.][]{wo95} shows that the pressure equilibrium can be 
maintained for H~{\sc i} in one of two stable ranges of kinetic temperature 
(T$_{\rm K}$), viz. $\sim 40$K $- 200$K for the CNM and $\sim 5000$K $- 8000$ K
for the WNM. H~{\sc i} at intermediate temperatures is unstable and is expected
to quickly migrate into one of the stable phases, unless energy is 
intermittently being injected into the medium. Recent observations 
and simulations indicate that in our Galaxy a significant fraction of 
the atomic ISM is in the thermally unstable region \citep[e.g.][]{va00,he01,
nref1,nref2}. 

The classical method of determining the temperature of the atomic
ISM is to compare the H~{\sc i} 21 cm line in absorption towards a bright 
continuum source with the emission spectrum along a nearby line of sight.
Assuming that the physical conditions are the same along both lines of
sight, one can measure the spin temperature (T$_{\rm S}$) (or excitation 
temperature) of the H~{\sc i} \citep[see e.g. ][for details]{ku88}. 
While the H~{\sc i} spin temperature, strictly speaking, characterizes 
the population  distribution between the two hyperfine levels of the 
hydrogen atom, it is often used as a proxy for the kinetic temperature 
of the gas. This is because, in high density regions, T$_{\rm S}$ is expected 
to be tightly coupled to the kinetic temperature via collisions, while in 
low density regions, resonant scattering of Lyman-$\alpha$ photons is 
generally expected to couple the spin temperature to the kinetic 
temperature \citep{fi58}. 

The 21 cm optical depth of the WNM is extremely low (typically $< 10^{-3}$)
which makes it very difficult to measure the H~{\sc i} absorption from
gas in the WNM phase. Consequently, emission-absorption studies usually 
provide only a lower limit to T$_{\rm S}$. If the particle and Lyman-$\alpha$ 
number densities are low, T$_{\rm S}$ could in turn be significantly 
lower than T$_{\rm K}$. On the other hand, one could use the 21 cm emission 
line width to determine an upper limit to the kinetic temperature. The
line width is an upper limit to the temperature because in addition
to thermal motions of the atoms, both bulk motion of the gas 
(e.g. differential rotation) as well as turbulence contribute to
the observed line width.

The presence of turbulence in the atomic ISM of our own Galaxy can be
detected through, for e.g. the scale free nature of the power spectrum
of the intensity fluctuations in H~{\sc i} 21 cm emission \citep{cro83,green93}.
In a turbulent medium, one would also expect the velocity dispersion to
increase as a power of the length scale. Such a power law velocity width
length scale scaling has been observed in the atomic ISM of the 
Large Magellanic Cloud (LMC) \citep{kim07}. To the best of our knowledge it 
has not been observed in the atomic ISM of our own Galaxy. In this paper we 
show that, assuming that the atomic ISM is in rough pressure equilibrium, the 
data from the millennium Arecibo 21 cm absorption-line survey 
\citep{ht03a, ht03b} is consistent with a velocity-length scale relation of 
the form $\sigma^2_v \propto l^{\alpha}$. We also show that this scaling is, 
to zeroth order, consistent with that expected from turbulence in an
medium with magnetic field of $\sim $ few $\mu$Gauss.

Once one has an estimate of the turbulent velocity contribution to the 
observed velocity width, one can correct for it, to derive a tighter limit to 
the kinetic temperature. We show that this correction leads to a substantially 
smaller fraction of the gas being in the WNM phase than if one does not take 
turbulence into account.

\section{Data, Analysis and Results}
\label{sec:dar}

\begin{figure}
\begin{center}
\includegraphics[scale=0.65, angle=0.0]{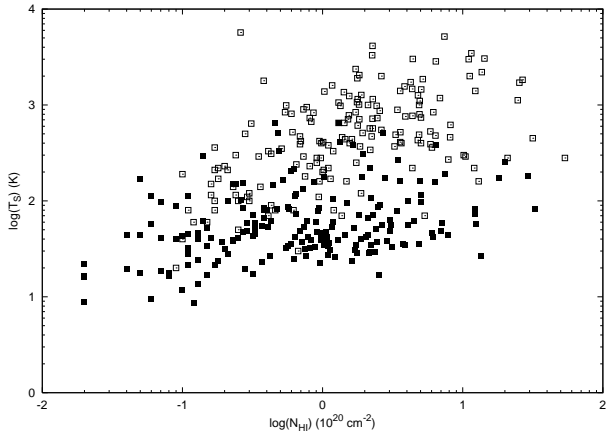}
\caption{\label{fig:1} \small{N$_{\rm HI}$ -- T$_{\rm S}$ plot for Gaussian 
components from the millennium Arecibo 21 cm absorption-line survey data. The 
filled squares are components detected both in emission and absorption and the 
empty squares are components detected only in emission giving a lower limit of 
T$_{\rm S}$.}}
\end{center}
\end{figure}

The data we use are taken from the millennium Arecibo 21 cm absorption-line 
survey and consist of emission and absorption spectra with a velocity 
resolution  of $\sim 0.4$ km s$^{-1}$ towards a total of 79 background radio 
sources. The observational and analysis techniques are discussed in detail by 
\citet{ht03a} and the astrophysical implications are discussed in 
\citet{ht03b}. A brief summary is that the absorption spectra were corrected
for emission from gas in the telescope beam by interpolating multiple
off-source spectra, after which both the emission and absorption spectra
were modeled as a collection of multiple Gaussian components. For each
component, the spin temperature, upper limits on kinetic temperature, 
column densities and velocities were derived using these fits. There are
several systematic uncertainties in such an analysis, discussed for
example in \citet{ht03a}, in particular those associated with estimating
and subtracting the emission, and the assumption that each Gaussian
component is a physically distinct entity. While a more robust measurement
of the absorption spectra can be done using interferometric observations
\citep[e.g.][]{ka03}, here we work with the fit parameters provided
as part of the survey. All the systematic uncertainties relevant to
the Arecibo millennium absorption survey hence also apply to our
results.

The survey lists a total of 374 Gaussian components in the emission spectra 
towards the 79 continuum sources Out of these, 205 components are also  
detected in H~{\sc i} absorption and have T$_{\rm S}$ measurements. For 21 of 
these components either the spin temperature had to be set to zero (by hand) 
in order to  attain convergence of the fit \citep{ht03a} or the upper limit of 
the kinetic  temperature computed from the line width is less than the spin 
temperature (or the lower limit of the spin temperature). The derived 
parameters for these components are clearly unphysical, and we do not use them 
in our analysis. We are hence left with a total of 353 Gaussian components 
consisting of 188 components detected both in emission and absorption and 165 
components detected only in emission. In Figure (\ref{fig:1}) is a scatter 
plot of the spin temperature T$_{\rm S}$ against the column density 
N$_{\rm HI}$; filled points are components that have been detected in both 
emission and absorption while empty points are components that have been 
detected only in emission. It is clear from the plot that the survey did not 
detect much gas with T$_{\rm S} \lesssim 10$ K.  

\begin{figure}
\begin{center}
\includegraphics[scale=0.65, angle=0.0]{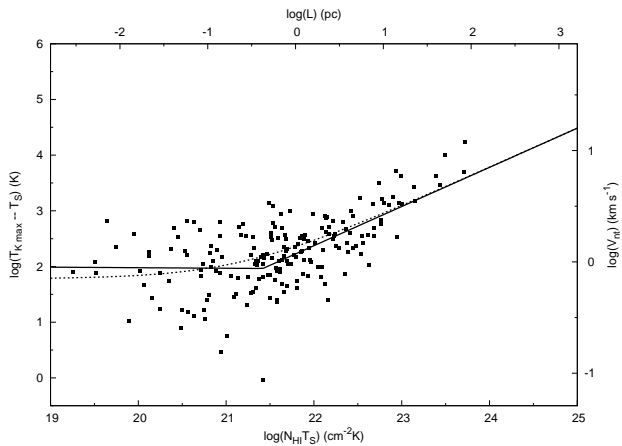}
\caption{\label{fig:2} \small{N$_{\rm HI}$T$_{\rm S}$ -- T$_{{\rm K}max} -$ 
T$_{\rm S}$ plot for Gaussian components from the millennium Arecibo 21 cm 
absorption-line survey data. Components detected both in emission and in 
absorption are only shown here. The solid line is a fit to the data using two 
power laws. The dotted line is the model with a $60$ K overestimation of 
T$_{{\rm K}max}$ (see \S\ref{sec:dar} for details).}}
\end{center}
\end{figure}

For a homogeneous cloud at temperature $T$, the pressure $P = nkT$, with
$n=$ N$_{\rm HI}/l$, where N$_{HI}$ is the column density and $l$ is the
length of the cloud. Putting these together we have $l =$N$_{\rm HI} kT/P$.
For the CNM clouds detected in absorption, it is quite reasonable to assume 
that the kinetic temperature is the same as the spin temperature T$_{\rm S}$.
If we further assume that the pressure is roughly constant across clouds,
then we have $l \propto $N$_{\rm HI}$\ T$_{\rm S}$. Though the density and 
temperature of neutral ISM vary over a few orders of magnitude, this 
assumption is justified because the pressure changes, in most of the cases, 
only by a factor of a few since the turbulence in the gas is at most transonic. 
Further, for these components, the  non-thermal component of the line width is 
given by  $v^2_{nt} \propto$ (T$_{{\rm K}max}$ $-$ T$_{\rm S}$), where 
T$_{{\rm K}max}$ is the measured line width of this component. In Figure 
(\ref{fig:2}) we show a scatter plot of (T$_{{\rm K}max}$ $-$ T$_{\rm S}$) 
against N$_{\rm HI}$T$_{\rm S}$, as discussed above, to zeroth order this can 
be regarded as a plot of non thermal velocity against length scale. The solid 
line in the figure is a dual power law fit; at large length scales 
(log(N$_{\rm HI}$T$_{\rm S}$) $\geq 21.4 \pm 0.2$) the power law index is 
$0.7 \pm 0.1$, while at small length scales, the power law index is consistent 
with zero. The dotted line shows a fit which assumes that the measured 
T$_{{\rm K}max}$ is larger than the true T$_{{\rm K}max}$ by 60K; it provides 
a reasonable fit to the data over five orders of magnitude in N$_{\rm HI}$\ 
T$_{\rm S}$. The length scale corresponding to a pressure of 2000 cm$^{-3}$K, 
as well as the non thermal velocity width in km/s are also indicated in the
figure. 

The correlation between cloud scale length and the non-thermal line width 
that we see at long length scales is consistent with a turbulence
driven velocity line-width relation. The assumptions we have made
(viz. homogeneous clouds, constant pressure) are fairly naive, and are
certainly not exactly valid in the current situation. The large scatter 
around the fit would be partly due to a break down of the assumptions
(e.g. variations in the pressure) and partly due to measurement errors.
If variations in pressure were the dominant contribution to the scatter,
one would expect to see a systematic variation of the scatter plot for
components in different Galactic latitude ranges. However, no such
variation is seen.

The ISM is known to have clumpy density and velocity structures and is believed 
to be turbulent at scales ranging from au to kpc \citep{di76,la81,de00}.
Incompressible hydrodynamic turbulence leads to the famous Kolmogorov scaling 
$\sigma^2_v \propto l^{2/3}$ \citep{ko41}, similar to what we see at large
length scales. However, the Galactic ISM cannot be modeled simply as an 
incompressible fluid. Recent theoretical studies and numerical simulations 
have investigated in details the turbulence of multi-phase medium \citep{nref10,
nref9,nref8,nref3,nref2}. In some of these cases\citep{nref3,nref2}, synthetic 
H~{\sc i} spectra are computed to study the effect of turbulence. In general, 
these analytical and numerical works predict a Kolmogorov-like turbulence in 
two-phase neutral ISM. \citet{nref2} also report, based on simulation results, 
a power law scaling $\sigma^2_v \propto l^{0.8}$ consistent with our 
observation. 

Now, since fractional ionization couples the H~{\sc i} to the magnetic field, 
the turbulence is expected to be magnetohydrodynamic (MHD) in nature. Though 
simple and ingenious models \citep[e.g.][]{gs95} of incompressible MHD 
turbulence have been proposed, most of the insights into incompressible and 
compressible MHD turbulence again come from numerical simulations 
\citep[][and references therein]{ch02}. Models \citep[like][]{gs95} predict a 
Kolmogorov-like energy spectrum, $E(k) \propto k^{-5/3}$, for incompressible 
MHD turbulence and this is supported by both numerical simulations and 
observations (see \citet{ch02} for details). In case of compressible MHD 
turbulence, Alfv$\acute{e}$n modes are least susceptible to damping mechanisms 
\citep{mi97} and hence the energy transfer in Alfv$\acute{e}$n waves is of 
major interest. Again, numerical simulations show that the energy spectra of 
Alfv$\acute{e}$n modes follow a Kolmogorov-like spectrum. 

In a situation where the bulk of the energy transfer is via Alfv$\acute{e}$n
waves, the non-thermal velocity dispersion $\delta v$ is related to the 
magnetic perturbation amplitude $\delta B$ and H~{\sc i} number density $n_H$ 
as $\delta v = \delta B/\sqrt{4\pi\mu n_{\rm H}m_{\rm H}}$ \citep{am75,ar07} 
where $\mu=1.4$ is the effective mass of an H+He gas with cosmic abundance, 
$m_{\rm H}$ is the mass of the hydrogen atom and it is usually assumed that 
$\delta B \sim B$. Using this relation, the magnetic field is found to be of 
the order of few $\mu$G (column density weighted mean and median values are 
11.7 and 10.2 $\mu$G respectively) with no significant trend related to 
``cloud'' size. We note that there are various uncertainties to the derived 
equipartition magnetic field. But our estimate is broadly consistent with the 
observed magnetic field in the diffuse neutral ISM and matches, within a 
factor of 2, with the median magnetic field estimated for a sub-sample of 
these components using Zeeman splitting measurements \citep{ht05}. 

The break that is clearly seen in Figure~(\ref{fig:2}) requires some 
attention. This change in the power law index can not be explained just in 
terms of lower  signal to noise on the physical quantities at low 
N$_{\rm HI}$T$_{\rm S}$ end. However, as shown in the figure, the data are 
well fit by a model in which the line width is overestimated by about $\sim 
60$ K. There are three systematic effects that may contribute to the 
overestimation of the line width without much affecting N$_{\rm HI}$ and 
T$_{\rm S}$: (i) the finite spectral resolution, (ii) blending of two or more 
narrow components and (iii) velocity (but not T$_{\rm S}$) fluctuations in the 
gas within the Arecibo beam. The contribution from the first effect is 
quantified by estimating the width of a Gaussian signal after smoothing it to 
a spectral resolution of 0.4 km s$^{-1}$ and adding noise similar to that in 
the actual spectra. The effect is found to be almost negligible because of the 
high spectral resolution. A similar numerical exercise with two Gaussian 
components was done to check the effect of blending of narrow components and 
ambiguities in Gaussian fitting. In this case, the effect is most significant
when the blended lines are of comparable amplitudes and have separations 
comparable to their widths. For example, blending of components with 
T$_{\rm K}max = 60$ K (width of the Gaussian $\sim 0.7$ km s$^{-1}$) with a 
separation of $\sim 1.2$ km s$^{-1}$ results in typically 20 -- 30 K 
overestimation of T$_{\rm K}max$. When the amplitudes of two Gaussian profiles 
are comparable, the line width is overestimated by upto $\sim 60$ K. The third 
possibility, that is, a fine scale structure in the velocity (but not in the 
temperature) has been proposed earlier \citep[e.g.][]{br05,nr06} to explain 
the observed fine scale H~{\sc i} opacity fluctuations \citep{di76,cr85}. Such 
velocity fluctuations within the Arecibo beam will also cause an 
overestimation of T$_{\rm K}max$. We, however, note that the scale length 
(inferred from N$_{\rm HI}$ and T$_{\rm S}$) of the components below the break 
is very small. Although the existence of tiny ``clouds'' is supported by 
observations and numerical simulations \citep[e.g.]{bk05,st05,na06,nref3, 
nref1}, their origin and physical properties are still unknown. The 
evaporation timescale for these clouds are $\sim 1$ Myr. These structures can 
survive if either the ambient pressure around the clouds is much higher than 
the standard ISM pressure or they are formed continuously with a comparable 
timescale. While we have presented plausible arguments for the break that we 
see not corresponding to a physical phenomena, the lack of detailed 
understanding of these tiny H~{\sc i} structures means that we can not rule 
out the possibility of some physical phenomenon being responsible for the 
break.

\subsection{A new indicator of the temperature}
\label{sec:ind}

\begin{figure}
\begin{center}
\includegraphics[scale=0.65, angle=0.0]{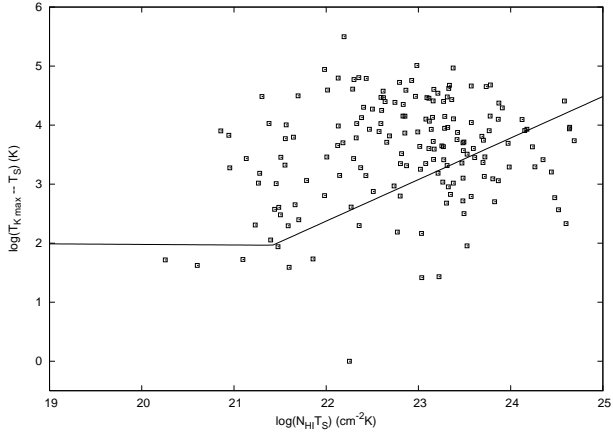}
\caption{\label{fig:3} \small{N$_{\rm HI}$T$_{\rm S}$ -- T$_{{\rm K}max} -$ 
T$_{\rm S}$ plot for Gaussian components from the millennium Arecibo 21 cm 
absorption-line survey data. Components detected only in emission and not in 
absorption are shown here and T$_{\rm S}$ is the lower limit for these 
components. The solid line is derived from the components detected both in 
emission and in absorption.}}
\end{center}
\end{figure}

\begin{figure}
\begin{center}
\includegraphics[scale=0.65, angle=0.0]{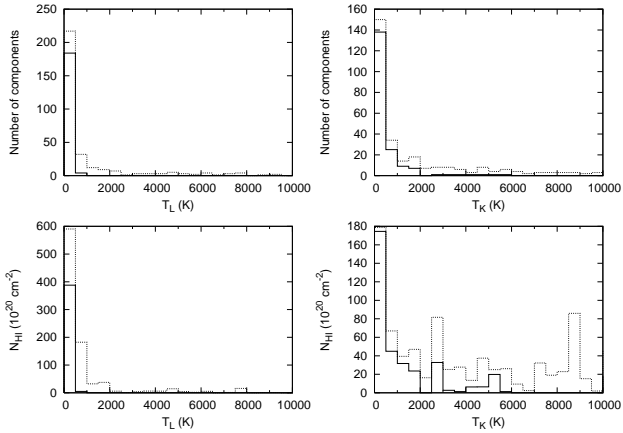}
\caption{\label{fig:4} \small{Histogram of derived temperatures T$_{\rm L}$ 
and T$_{\rm K}$. The solid lines only include components detected both in 
emission and in absorption and the dotted lines include all the components. 
The top two panels give the number of Gaussian components and the bottom 
panels give the H~{\sc i} column density. The left two panels are histogram of 
T$_{\rm L}$ and the right two are of T$_{\rm K}max$.}}
\end{center}
\end{figure}

\begin{figure}
\begin{center}
\includegraphics[scale=0.65, angle=0.0]{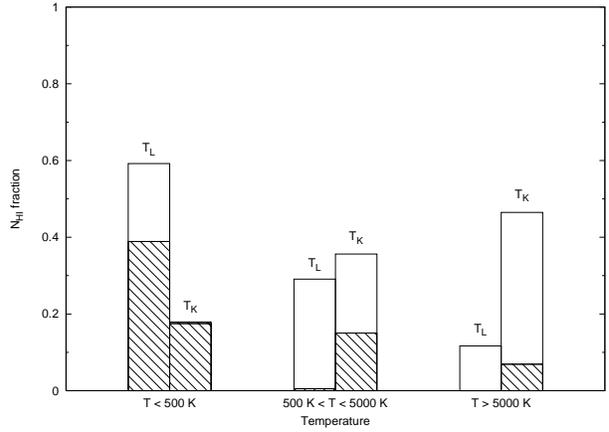}
\caption{\label{fig:4a} \small{N$_{\rm HI}$ fraction, for components detected 
both in emission and in absorption (hatched histogram) and components detected 
only in emission and not in absorption (empty histogram), in different 
temperature range using T$_{\rm L}$ and T$_{{\rm K}max}$ as the proxy for the 
actual physical temperature.}}
\end{center}
\end{figure}

For a multi-phase medium if the turbulent velocity dispersion scaling is 
similar for coexisting phases, then this scaling relation can be exploited to 
get a handle on the physical temperature of the gas that is detected only in 
H~{\sc i} emission but not in absorption. Since one only has a lower limit
on  T$_{\rm S}$ for these components, they lie, as expected, systematically 
on the top left side of the fit to the components detected in both
emission and absorption (Figure (\ref{fig:3})). For these components 
we define a proxy temperature T$_{\rm L}$ that will restore the component 
back to this power law correlation. Given the measured N$_{\rm HI}$ and 
T$_{{\rm K}max}$ from the emission spectra, one can uniquely compute this 
proxy temperature. Since T$_{\rm L}$ corresponds to the velocity width after
correction for the turbulent velocity, it is a better estimate of the actual 
physical temperature of the cloud than that of T$_{{\rm K}max}$.
Note that since most of the components in Figure~(\ref{fig:3}) line beyond
the break in the fitted function  the derived T$_{\rm L}$ is independent
of the whether the break arises due to some underlying physical reason.

For all except 2 of the 165 components detected only in emission, T$_{\rm L}$ 
was calculated as described above. For two components, no meaningful solution 
for T$_{\rm L}$ could be found, and they are hence not included in the further 
analysis. For the components detected both in emission and in absorption, 
T$_{\rm S}$ is taken to be same as T$_{\rm L}$. With this, we have T$_{\rm L}$ 
and T$_{{\rm K}max}$ for a total of 351 Gaussian components. Figure 
(\ref{fig:4}) shows the histogram of T$_{\rm L}$ and T$_{{\rm K}max}$ in terms 
of both number of ``clouds'' and H~{\sc i} column density. The top two panels 
give the number of Gaussian components and the bottom panels give the 
H~{\sc i} column density in different temperature bin. From the histograms, it 
is evident that our results qualitatively confirm the earlier detection of a 
significant fraction of gas in the thermally unstable region. Quantitatively, 
however, a significant fraction of the gas with high T$_{{\rm K}max}$ after 
correction for turbulent broadening corresponds to gas in the stable phase. 
This quantitative difference is illustrated in Figure (\ref{fig:4a}) which 
shows the N$_{\rm HI}$ fraction for both the population (components detected 
both in emission and in absorption and components detected only in emission) 
in different temperature range using T$_{\rm L}$ instead of T$_{{\rm K}max}$ 
as the proxy for actual physical temperature. Further, while \citet{ht03b} 
find that $60\%$ of all H~{\sc i} is in the WNM phase, we find that only 
$40\%$ of the gas has temperature higher than $500$ K.  

\begin{figure}
\begin{center}
\includegraphics[scale=0.65, angle=0.0]{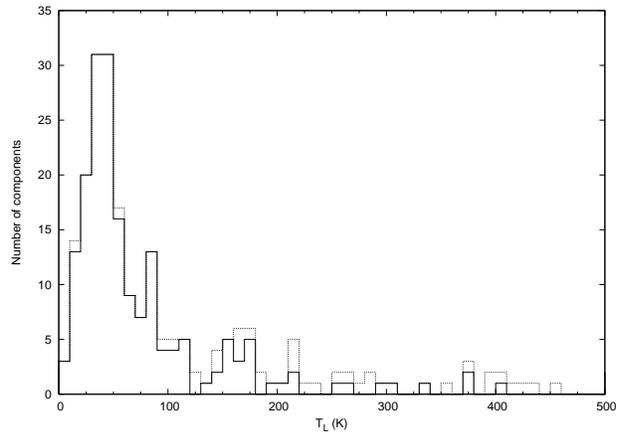}
\caption{\label{fig:5} \small{Histogram of derived temperatures for T$_{\rm L} 
< 500$ K. The solid lines only include components detected both in emission 
and in absorption and the dotted lines include all the components.}}
\end{center}
\end{figure}

\begin{figure}
\begin{center}
\includegraphics[scale=0.65, angle=0.0]{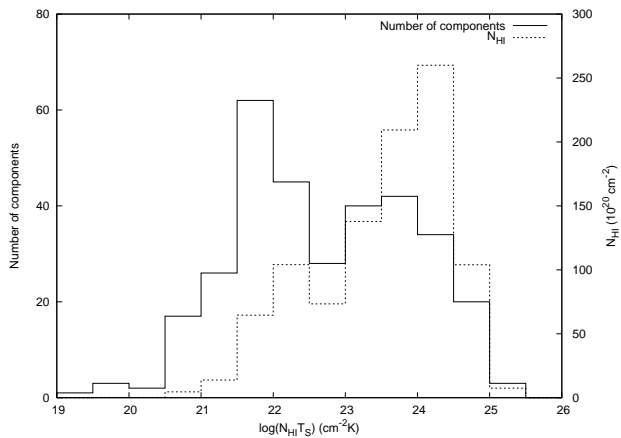}
\caption{\label{fig:6} \small{Histogram of scale length $L \sim$ N$_{\rm HI}$ 
T$_{\rm S}$ in terms of number of components (solid line) and total H~{\sc i} 
column density (dotted line). Two distinct peaks are evident in both cases.}}
\end{center}
\end{figure}

A closer examination of the components with T$_{\rm L} < 500$ shows a clear 
peak near T$_{\rm L} \sim 50$ K in the number distribution and that the major 
fraction of the gas is below T$_{\rm L} \sim 100$ K as shown in Figure 
(\ref{fig:5}). Figure (\ref{fig:6}) shows the histogram of scale length $L 
\sim$ N$_{\rm HI}$ T$_{\rm S}$. This clearly shows a bi-modal statistical 
distribution of the ``cloud'' size for the neutral ISM. As expected, the 
dominant contribution to the the peak at lower $L$ is from the cold components 
and to the peak at higher $L$ is mostly from the warm components. If $N$ is 
number of clouds along the lines of sight, $R$ is typical size of the clouds 
and $n$ is H~{\sc i} number density, then N(H~{\sc i})$_X \propto N_X R_X n_X$ 
where $X$ stands for CNM or WNM. From the observed N(H~{\sc i}) and $N$ for 
all the components used for this analysis we find, using this relation, 
$R_{\rm WNM}/R_{\rm CNM} \sim 110$ for typical $n_{\rm CNM}/n_{\rm WNM} \sim 
100$. This is consistent with the ratio of the length scale corresponding to 
two peaks in the bi-modal distribution.

\section{Conclusions}
\label{sec:con}

In this work we present a new phenomenology based technique to address the 
issue of non-thermal line width and the temperature of the diffuse neutral 
hydrogen of our Galaxy assuming a rough pressure equilibrium between different 
phases of the ISM. A possible connection between the observed Kolmogorov-like 
scaling of the non-thermal velocity dispersion in the Galactic H~{\sc i} and 
the turbulence of the interstellar medium is discussed. This scaling relation 
is used to re-examine the issue of the temperature of the Galactic ISM with 
the help of the millennium Arecibo 21 cm absorption-line survey measurements. 
The distribution of the derived temperature is found to be significantly 
different from the distribution of the upper limits of the kinetic 
temperature. A considerable fraction ($\sim 29\%$) of the gas is found to be 
in the thermally unstable phase, qualitatively confirming earlier results. 
However, about $60\%$ of all the neutral diffuse gas, a much higher fraction 
than that of reported earlier, has temperature below $500$ K. The CNM 
temperature distribution shows a clear peak near T $ \sim 50$ K and the cloud 
size for the neutral ISM shows a bi-modal statistical distribution. Derived 
magnetic field from the non-thermal velocity dispersion matches, within a 
factor of 2, with the magnetic field value estimated from the Zeeman splitting 
measurements. The Kolmogorov-like scaling is consistent with the existing 
theoretical prediction, numerical simulations and earlier observational 
results.  

\section*{Acknowledgements}

This research has made use of the data from the millennium Arecibo 21 cm 
absorption-line survey measurements and NASA's Astrophysics Data System. We 
are grateful to Rajaram Nityananda and Nissim Kanekar for their comments on an 
earlier version of the Letter. We thank K. Subramanian, D. J. Saikia and R. 
Srianand for useful discussions. One of the authors (LP) would like to 
acknowledge the hospitality of all the staff members of the National Centre 
for Radio Astrophysics (NCRA) during her stay for the Visiting Student 
Research Programme (2007). We are grateful to the anonymous referee for 
prompting us into substantially improving this paper. This research was 
supported by the National Centre for Radio Astrophysics of the Tata Institute 
of Fundamental Research (TIFR).

\bsp

\label{lastpage}

\end{document}